\documentclass[aip,jap,reprint, floatfix]{revtex4-1}
\usepackage[pdftex]{graphicx}
\usepackage{bm}
\usepackage[range-units = single]{siunitx}
\usepackage{amsmath}
\usepackage{amsfonts}
\usepackage{multirow}
\usepackage{longtable}
\usepackage{wasysym}
\usepackage[colorlinks=true, allcolors=blue]{hyperref}

\newcommand{\vect}[1]{\bm{#1}}

\newcommand*\diff{\mathop{}\!\mathrm{d}}
\newcommand{\norm}[1]{\left\lVert #1 \right\rVert}
\newcommand{\const}[1]{\operatorname{#1}}
\newcommand{\ber}[1]{p_\mathrm{err}^\mathrm{#1}}

\DeclareMathOperator\erf{erf}

\begin{document}

\title{Switching field distribution of exchange coupled ferri-/ferromagnetic composite bit patterned media}
\author{Harald Oezelt}
\email{harald.oezelt@donau-uni.ac.at}
\thanks{The following article has been submitted to the Journal of Applied Physics. After it is published it will be found at \url{http://jap.aip.org}. This article may be downloaded for personal use only. Any other use requires prior permission of the author and AIP Publishing. Copyright (2015) American Institute of Physics.}
\author{Alexander Kovacs}
\author{Johann Fischbacher}
\affiliation{Center for Integrated Sensor Systems, Danube University Krems, 2700 Wiener Neustadt, Austria}
\author{Patrick Matthes}
\affiliation{Institute of Physics, University of Augsburg, 86159 Augsburg, Germany}
\author{Eugenie Kirk}
\author{Phillip Wohlh\"uter}
\author{Laura Jane Heyderman}
\affiliation{Laboratory for Mesoscopic Systems, Department of Materials, ETH Zurich, 8093 Zurich, Switzerland}
\affiliation{Laboratory for Micro- and Nanotechnology, Paul Scherrer Institute, 5232 Villigen PSI, Switzerland}
\author{Manfred Albrecht}
\affiliation{Institute of Physics, University of Augsburg, 86159 Augsburg, Germany}
\author{Thomas Schrefl}
\affiliation{Center for Integrated Sensor Systems, Danube University Krems, 2700 Wiener Neustadt, Austria}

\begin{abstract}
We investigate the switching field distribution and the resulting bit error rate of exchange coupled ferri-/ferromagnetic bilayer island arrays by micromagnetic simulations. Using islands with varying microstructure and anisotropic properties, the intrinsic switching field distribution is computed. The dipolar contribution to the switching field distribution is obtained separately by using a model of a triangular patterned island array resembling $\SI{1.4}{Tb/in^2}$ bit patterned media. Both contributions are computed for different thickness of the soft exchange coupled ferrimagnet and also for ferromagnetic single phase FePt islands. A bit patterned media with a bilayer structure of FeGd($\SI{5}{nm}$)/FePt($\SI{5}{nm}$) shows a bit error rate of $10^{-4}$ with a write field of $\SI{1.16}{T}$.
\end{abstract}


\maketitle

\section{Introduction}
The concept of bit patterned media (BPM) is one of several promising approaches to push data density in magnetic storage devices beyond the limits of conventional perpendicular magnetic recording (PMR)~\cite{Terris2006, Piramanayagam2009, Albrecht2015}. In this scheme the recording media is an array of decoupled magnetic single domain islands where each dot stores one data bit. While this approach reduces bit transition jitter and improves signal to noise ratio (SNR) compared to PMR~\cite{Albrecht2002}, the writeability and thermal stability of the magnetic islands need to be maintained. This can be addressed by designing the islands as exchange spring or exchange coupled composite (ECC) structures which are made of at least two layers of different magnetic anisotropy~\cite{Victora2002, Suess2007, Goncharov2007}. In such islands a magnetically hard layer (usually FePt) ensures the thermal stability and an exchange coupled softer layer is used to decrease the required writing field~\cite{Sbiaa2009}. Ferrimagnetic (FI) materials such as FeGd or FeTb might serve as ideal candidates for the soft layer since their magnetic properties can be tailored easily by their composition. Also in combination with the concept of heat assisted magnetic recording (HAMR), such materials bear potential when exploiting their compensation point.

One of the challenges on the way towards BPM is to reduce the switching field distribution (SFD) of the island array to ensure a low bit error rate (BER)~\cite{Richter2006}. The SFD origins of islands can be categorized in intrinsic contributions and stray-field contributions. While the latter deals with the dipolar field interactions between neighbouring islands~\cite{Hellwig2007, Pfau2014,Repain2004}, the intrinsic part stems from the variation of magnetic properties, shape and size of the islands~\cite{Thomson2006, Pfau2011, Lee2011}. Obviously the development of suitable fabrication processes for BPM plays an important role in improving the intrinsic SFD. 
In this paper, however, we investigate how an exchange-coupled fer\-ri\-ma\-gne\-tic layer influences the SFD. Due to the changed magnetization reversal process of exchange spring media and also the amorphous structure of fer\-ri\-ma\-gne\-tic materials, this crucial property might be improved significantly~\cite{Kikitsu2009, Krone2011}. We compute the distributions of the intrinsic and dipolar interaction field contribution separately. To quantify the effect of both contributions, we compute the bit error rate of the medium but neglect other contributions like the head field gradient.

In the following, we briefly present the used micromagnetic model for such bilayer islands and start our investigation by looking at different island diameters and FI-layer thicknesses. We then compute the intrinsic SFD of the bilayer islands by varying the magnetocrystalline anisotropic properties and microstructure of both layers. The dipolar contribution to the SFD is then determined by computing the dipolar interaction field of an array of such islands. Finally we present the results for both contributions and show the influence of the dipolar interaction field on the BER for different island designs.

\section{Micromagnetic Simulations}
Our finite element calculations for the SFD are based on our previously presented model of ex\-chan\-ge-coupled fer\-ri-/fer\-ro\-ma\-gne\-tic he\-tero\-struc\-tu\-res~\cite{Oezelt20151, Oezelt20152}. Therefore, only a brief summary follows. To calculate the magnetization reversal of the magnetic islands, we use the common Landau-Lifshitz-Gilbert (LLG) equation for ferromagnets (FM) and an adapted LLG equation~\citep{Mansuripur95} for the ferrimagnetic (FI) layer. The latter is based on the assumption that in thin films the two sub-lattices of the FI are strongly coupled antiparallel and therefore their magnetic moments can be substituted by an effective net moment. This allows the combination of the LLG equation of both sub-lattices to obtain an effective equation for thin FI-layers. For all our simulations the damping constant is set to $\alpha_\mathrm{eff}=1$, since we are only interested in the static hysteresis to extract the switching field $H_\mathrm{sw}$.

In the following Section~\ref{sec:ber} we describe how the BER is computed by considering both the intrinsic and the dipolar interaction contribution. Then the respective simulations are described for computing the intrinsic SFD in Section~\ref{sec:intSFD} and computing the dipolar contribution to the SFD in Section~\ref{sec:dipSFD}. All simulations were performed using the finite element micromagnetic package FEMME~\cite{Schrefl2007}.

\subsection{\label{sec:ber}Bit Error Rate of bit patterned media}
\begin{figure}[htb]
\centering
\def\svgwidth{\linewidth}
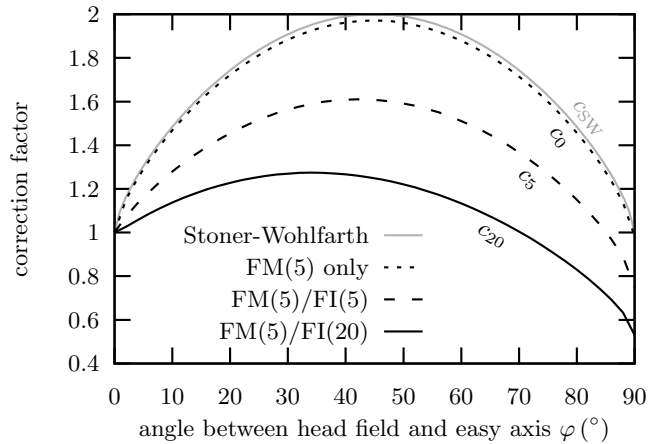
\caption{\label{fig:fcorr}Correction functions for the angular variation of the switching field. The gray solid curve is the evaluated Stoner-Wohlfarth function $c_\mathrm{SW}$ in \eqref{eq:swf}. The functions labelled according to the FI-layer thickness of the different island designs $c_{0}$, $c_{5}$ and $c_{20}$ are calculated using our micromagnetic model.}
\end{figure}

The intrinsic switching field distribution of the islands can be described with a Gaussian fit, with the mean switching field $\bar{H}_\mathrm{sw}$ and the standard deviation $\sigma_\mathrm{int}$:
\begin{align}
f(h) = \frac{1}{\sigma_\mathrm{int}\sqrt{2\pi}}\const{e}^{-\left(\frac{h-\bar{H}_\mathrm{sw}}{\sigma_\mathrm{int}\sqrt{2}}\right)^2}.
\end{align}
To switch the islands, an external head field $\vect{H}_\mathrm{head}$ is turned on. This field is applied either parallel or tilted with respect to the out-of-plane axis of the medium. For the latter the head field has to be corrected due to the angular variation of the switching field. For single phase islands an effective head field $H_\mathrm{head}^*$ can be calculated by using the Stoner-Wohlfarth model~\cite{Shukh2004,Stoner1948}. Hereby $\vect{H}_\mathrm{head}$ is scaled by a correction function $c$ depending on the angle $\varphi$ between the applied head field and the anisotropic easy axis of the islands (out-of-plane): 
\begin{align}
c_\mathrm{SW}(\varphi)=\left(\sin^{2/3}\varphi+\cos^{2/3}\varphi\right)^{3/2}.\label{eq:swf}
\end{align}
Using our micromagnetic model we calculate the switching field of different bilayer systems at different angles $\varphi$ and also express this angular dependence as correction functions. Later the island designs we investigate will mainly differ in the thickness of the ferrimagnetic (FI) layer $t_\mathrm{FI}$, and therefore their corresponding correction function are labelled as $c_{t_\mathrm{FI}}$. FIG.~\ref{fig:fcorr} shows the analytic Stoner-Wohlfarth correction function $c_\mathrm{SW}$ in gray and the simulated correction functions of three island designs with a diameter of $d=\SI{20}{nm}$ and a FI-layer thickness of $t_\mathrm{FI}=\SI{0}{nm}$, $\SI{5}{nm}$ and $\SI{20}{nm}$.
The effective head field is calculated by scaling the magnitude of $\vect{H}_\mathrm{head}$ with the respective correction function.
\begin{align}
H_\mathrm{head}^*=c_{t_\mathrm{FI}}(\varphi)\norm{\vect{H}_\mathrm{head}}\label{eq:hheff}
\end{align}

The numerically calculated correction function for the single ferromagnetic layer $c_0$ is very close to the theoretical $c_\mathrm{SW}$. When adding the exchange coupled FI-layer the correction factor is generally reduced and even below $1$ at higher angles. This means the switching field is less reduced by a tilted head field compared to single phase media or even increases at higher tilting angles. For bilayer islands the maxima of the correction functions move to lower $\varphi$. In other words, for soft-hard bilayer structures the minimum switching field resides at lower angles than for single phase islands. 

\begin{figure}[htb]
\centering
\def\svgwidth{\linewidth}
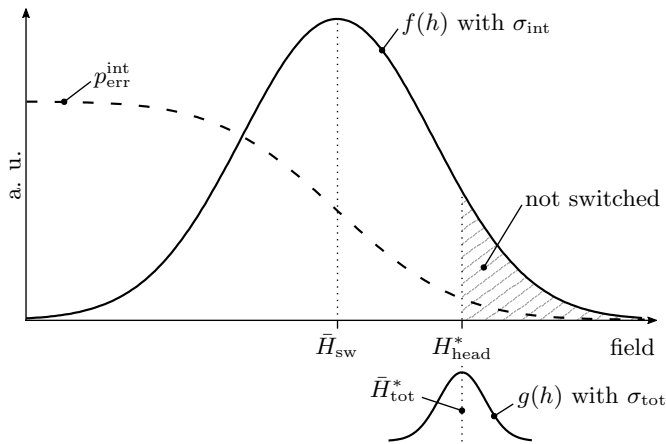
\caption{\label{fig:prop}Schematic of the bit error rate calculation. If the dipolar field is neglected the intrinsic switching field distribution (solid line) and the applied field $H_\mathrm{head}^*$ determine the bit error rate $\ber{int}$ (dashed line) according to \eqref{eq:perr}, since islands with $H_\mathrm{sw}>H_\mathrm{head}^*$ can not be switched (hatched area). When also taking into account the dipolar field, $H_\mathrm{tot}^*$ is applied with the distribution of the dipolar field $g(h)$.}
\end{figure}

With the respective effective head field as upper limit the probability of switching the islands is given by~\cite{Muraoka2011}:
\begin{align}
p_\mathrm{sw} &= \frac{1}{\sigma_\mathrm{int}\sqrt{2\pi}}\int\limits^{H_\mathrm{head}^*}_{-\infty}\const{e}^{-\left(\frac{h-\bar{H}_\mathrm{sw}}{\sigma_\mathrm{int}\sqrt{2}}\right)^2}\diff h\\
&=\frac{1}{2}\left(1+\erf\left(\frac{H_\mathrm{head}^*-\bar{H}_\mathrm{sw}}{\sigma_\mathrm{int}\sqrt{2}} \right)\right).
\end{align}
Therefore, the bit error rate for a set of isolated islands with an intrinsic switching field distribution of $\sigma_\mathrm{int}$ is $\ber{int}=1-p_\mathrm{sw}$ or
\begin{align}
\ber{int}=\frac{1}{2}\left(1-\erf\left(\frac{H_\mathrm{head}^*-\bar{H}_\mathrm{sw}}{\sigma_\mathrm{int}\sqrt{2}}\right)\right).\label{eq:perr}
\end{align}
The diagram in Fig.~\ref{fig:prop} shows the intrinsic distribution $f(h)$ (solid line) and the applied effective field $H_\mathrm{head}^*$. Islands with a switching field $H_\mathrm{sw}>H_\mathrm{head}^*$ (hatched area) cannot be switched and cause bit errors. The corresponding bit error rate $\ber{int}$ is shown as a dashed line.

However, in arrays of nano islands with increased areal density the magnetostatic dipole interaction between the islands has to be taken into account\cite{Muraoka2008}. Hence, the field acting on an island is the sum of the current head field and the dipolar interaction field exerted by the neighbouring dots: $\vect{H}_\mathrm{tot}=\vect{H}_\mathrm{head}+\vect{H}_\mathrm{dip}$. Again, either the perpendicular component of the total field can be used~\cite{Dong2011}, or it can be corrected by the angular variation of the switching field~\cite{Richter2006,Greaves2008} obtaining a total effective field: 
\begin{align}
H_\mathrm{tot}^*=c_{t_\mathrm{FI}}(\varphi)\norm{\vect{H}_\mathrm{head}+\vect{H}_\mathrm{dip}}\label{eq:hteff}
\end{align}
Here $\varphi$ is the angle between the total field $\vect{H}_\mathrm{tot}$ and the anisotropic easy axis of the islands. Because the dipolar interaction field is incorporated, the total effective field $H_\mathrm{tot}^*$ has the distribution of $\vect{H}_\mathrm{dip}$ imprinted. The $\vect{H}_\mathrm{dip}$ acting on an island is determined by the magnetic configuration of the neighbouring islands. By computing $\vect{H}_\mathrm{dip}$ for different randomized magnetic configurations, as described later in section~\ref{sec:dipSFD}, and then calculating $H_\mathrm{tot}^*$ for each configuration, we get the distribution of the total effective field. This distribution is again approximated by a Gaussian fit:
\begin{align}
g(h) = \frac{1}{\sigma_\mathrm{tot}\sqrt{2\pi}}\const{e}^{-\left(\frac{h-\bar{H}_\mathrm{tot}^*}{\sigma_\mathrm{tot}\sqrt{2}}\right)^2}.
\end{align}
Now the previously applied effective field $H_\mathrm{head}^*$ is substituted by a mean total field $\bar{H}_\mathrm{tot}^*$ carrying the distribution $g(h)$ (see Fig.~\ref{fig:prop}).
By integrating over the product of the total effective distribution $g\left(h\right)$ and the bit error rate of isolated islands $\ber{int}$, the overall bit error rate $\ber{tot}$ can be computed.
\begin{align}
\ber{tot}=\int\limits^{\infty}_{-\infty}\frac{g\left(h\right)}{2}\left( 1-\erf\left(\frac{h-\bar{H}_\mathrm{sw}}{\sigma_\mathrm{int}\sqrt{2}}\right)\right)\diff h\label{eq:perrTot}
\end{align}

In this work, especially in section \ref{sec:berRes}, we compare these two different approaches to compute the bit error rate:
\begin{itemize}
\item The dipolar interaction field is neglected and therefore an array of isolated islands is assumed. We apply an effective field \eqref{eq:hheff} and calculate $\ber{int}$ according to \eqref{eq:perr}.
\item The dipolar interaction field is incorporated by imprinting its distribution onto the effective field using \eqref{eq:hteff}. The total bit error rate $\ber{tot}$ is then calculated according to \eqref{eq:perrTot}.
\end{itemize}
In order to obtain the intrinsic distribution and the dipolar interaction field distribution, for both contributions a micromagnetic simulation scheme had to be set up. These two schemes and the used parameters are described in the following subsections \ref{sec:intSFD} and \ref{sec:dipSFD}.

\subsection{\label{sec:intSFD}Intrinsic switching field distribution}
We designed the islands as cylindrical dots with an $L1_0$ chemically ordered $\mathrm{Fe}_{52}\mathrm{Pt}_{48}$ layer as the ferromagnetic (FM) layer and an exchange coupled $\mathrm{Fe}_{74}\mathrm{Gd}_{26}$ layer as the ferrimagnetic (FI) layer. In this arrangement the FM-layer represents the magnetically hard phase which stores the bit while the FI-layer is the magnetically soft phase which helps to lower the switching field and reduce the SFD.

The $\SI{5}{nm}$-thick FM-layer has a granular structure with a average grain diameter of $\SI{14}{nm}$. Each grain exhibits its own random anisotropy constant and uniaxial anisotropy direction. The easy axes of the grains are within a cone angle of $\theta_\mathrm{max}=\SI{15}{\degree}$. The cone angle defines the maximum deviation of the magneto-crystalline anisotropy axis from the film normal ($z$-axis). Although the FI-layer is amorphous, we incorporate material inhomogeneities by dividing the layer into patches~\cite{Mansuripur1991} with an average size of $~\SI{10}{nm}$. Again, the anisotropic properties vary across the patches, but for the FI-layer the uniaxial direction is not confined to a certain cone angle. The thickness of the FI-layer $t_\mathrm{FI}$ is varied between $\SI{3}{nm}$ and $\SI{50}{nm}$ in our simulations. A representation of a bilayer dot with a granular FM-layer and inhomogeneous FI-layer is shown as an inset in Fig.~\ref{fig:sfdotsize}. The tessellation of both layers into grains and patches, was generated with the software Neper~\cite{Quey2011}.

Since the two layers possess a different microstructure, they are meshed separately. The layers are exchange coupled by calculating the interface exchange energy and incorporating it into the effective field equation of the respective layer. The coupling energy of $J_\mathrm{ixhg}=\SI{5}{mJ\per\m^2}$ between the two layers was chosen to match the coercive field of experimental data. 
The exchange constant $A_\mathrm{x}$, the mean anisotropic constant $\bar{K}_\mathrm{u}$, the standard deviation of the anisotropy constant $\sigma_K$, the cone angle of the easy axes $\theta_\mathrm{max}$ and the saturation polarisation $J_\mathrm{s}$ of both layers are listed in TABLE~\ref{tab:properties}. These values are extracted from experiments and fitting computed magnetization reversal curves to experimentally measured curves. For the material parameters the exchange length $l_\mathrm{x}=\sqrt{\mu_0A_\mathrm{x}/J_\mathrm{s}^2}$ and the Bloch wall parameter $\delta_0=\sqrt{A_\mathrm{x}/K_\mathrm{u}}$ are $\SI{2.8}{nm}$ and $\SI{3.2}{nm}$ for FePt and $\SI{5.9}{nm}$ and $\SI{16}{nm}$ for FeGd. With a mesh size which is kept always smaller than $\SI{2.5}{nm}$ we hope to represent domain walls correctly.

\renewcommand\arraystretch{1.2}
\begin{table}[htb]
\centering
\caption{\label{tab:properties}Intrinsic magnetic properties of fer\-ri-/fer\-ro\-ma\-gne\-tic (FI/FM) bilayer islands extracted from experiments.}
\begin{ruledtabular}
\begin{tabular}{@{\extracolsep{\fill}}crrrrrr}
\multirow{2}{*}{layer}
& \multicolumn{1}{c}{$A_\mathrm{x}$} & \multicolumn{1}{c}{$\bar{K}_\mathrm{u}$} & \multicolumn{1}{c}{$\sigma_K$} & \multicolumn{1}{c}{$\theta_\mathrm{max}$} & \multicolumn{1}{c}{$J_\mathrm{s}$} \\
& \multicolumn{1}{c}{$\left(\si{pJ/m}\right)$} & \multicolumn{1}{c}{$\left(\si{kJ/m^3}\right)$} & \multicolumn{1}{c}{$\left(\si{\%}\right)$} & \multicolumn{1}{c}{$\left(\si{\degree}\right)$} & \multicolumn{1}{c}{$\left(\si{mT}\right)$} \\
\hline
FM (FePt) & 10  & 975 & 5   & 15 & 1257 \\
FI (FeGd) & 2   &   8 & 20  & 90 &  268 \\
\end{tabular}
\end{ruledtabular}
\end{table}

The reversal curve of a bilayer island is computed by saturating the island with an applied external field $H_\mathrm{ext}$ in the positive out-of-plain direction $z$ and then decreasing the field in the opposite direction until the island is switched. The switching field $H_\mathrm{sw}$ is then obtained by extracting the value of the external field when half of the hard FM-layer is switched,  $M_z^\mathrm{FM}\left(H_\mathrm{sw}\right)=0$. To get the intrinsic SFD we evaluate the reversal curves of one hundred islands. Each island differs only in the randomly generated microstructure and its anisotropic properties within the limits described above. Therefore, other contributions like variation in island diameter or damage due to patterning are not considered. 

\subsection{\label{sec:dipSFD}Dipolar interaction field contribution}
When looking at the dipolar interaction field we neglect the microstructure of both layers and model an array of equivalent islands, still consisting of two exchange coupled layers, but both layers with perfect out-of-plane anisotropic easy axis. The intrinsic properties $A_\mathrm{x}$, $\bar{K}_\mathrm{u}$ and $J_\mathrm{s}$ are kept as defined in TABLE~\ref{tab:properties}. The array consists of $11\times11$ islands where, at the center, one island is left out. This gap marks the position where the dipolar interaction field is measured. The dipolar field of these 120 islands is calculated by letting the system relax after randomly assigning an initial magnetization to each island. 

\begin{figure}[htb]
\centering
\def\svgwidth{\linewidth}
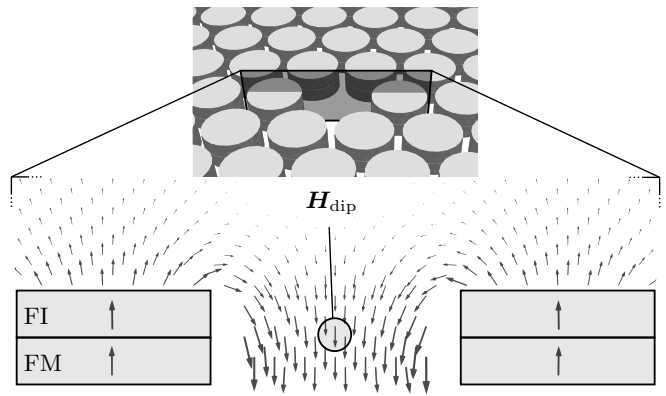
\caption{\label{fig:diSchematic}Schematic of the geometric simulation model to determine the dipolar interaction field acting on an island in an array of bilayer islands.}
\end{figure}

To compile a histogram of the interaction field $\vect{H}_\mathrm{dip}$ acting on an island we repeat this simulation 500 times, assigning each time a new random initial magnetization, and determine the interaction field $\vect{H}_\mathrm{dip}$ at the center of the virtual island (the gap) at the position of the interface between the ferromagnetic and the ferrimagnetic layer. This choice of position is justified by the reversal mechanism in ECC media, where the critical process is the expansion of a reversed domain into the hard magnetic phase. In our bilayer structure, the ferromagnet (FePt) acts as hard magnetic phase and the ferrimagnet (FeGd) acts as soft phase. 
The simulation arrangement is illustrated in Fig.~\ref{fig:diSchematic} with two neighbouring islands, their exerted stray field and the probe point in the gap, where $\vect{H}_\mathrm{dip}$ is determined. 

Of course the dipolar interaction field is strongly dependent on the density and pattern of the array. In this work we investigate an array of islands that are distributed in a triangular pattern. The diameter of the islands is chosen to $\SI{20}{nm}$. The triangular patterned array resembles BPM with $~\SI{1.4}{Tb/in^2}$ and has a x-pitch of $\SI{19.5}{nm}$ and a y-pitch of $\SI{22.5}{nm}$.

\section{Results and Discussion}
To obtain the switching field distribution and the bit error rate of the triangular island arrays, we start by calculating the intrinsic switching field and its distribution in FePt islands and FePt/FeGd bilayer islands. In this island configuration, the FePt layer is the magnetically hard ferromagnet (FM) and the FeGd layer is the soft ferrimagnet (FI). For further investigations we then limit our island configurations to three designs, all with a diameter of $d=\SI{20}{nm}$: 
\begin{itemize}
\item FM(5) only: single phase FePt islands with a thickness of $t_\mathrm{FM}=\SI{5}{nm}$.
\item FM(5)/FI(5): bilayer islands with equally thick hard and soft phase $t_\mathrm{FM}=t_\mathrm{FI}=\SI{5}{nm}$.
\item FM(5)/FI(20): bilayer islands with a FePt layer of $t_\mathrm{FM}=\SI{5}{nm}$ thickness and a soft FeGd phase with $t_\mathrm{FI}=\SI{20}{nm}$.
\end{itemize}
With these three configurations, we can compare single phase and exchange coupled composite islands, and also look at an increased FI-layer thickness.

\subsection{Intrinsic switching field distribution}
\begin{figure}[htb]
\centering
\def\svgwidth{\linewidth}
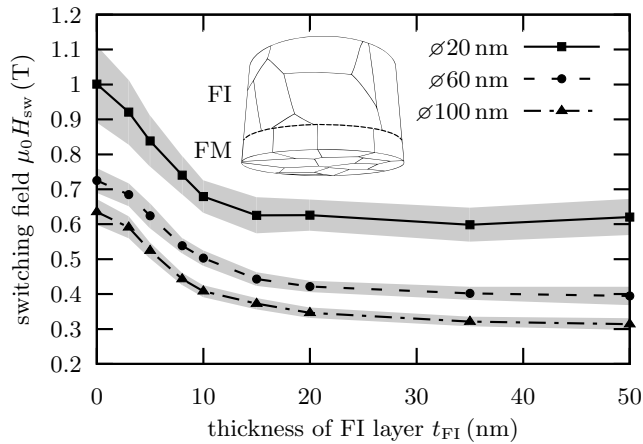
\caption{\label{fig:sfdotsize}Mean switching field of bilayer islands depending on the thickness of the FI soft layer for different island diameters ($\SI{20}{nm}$, $\SI{60}{nm}$ and $\SI{100}{nm}$). Standard deviation of the SFD is depicted as gray shading. The inset shows a typical geometric model of a FM/FI bilayer structure with patches in the FI- and columnar grains in the FM-layer.}
\end{figure}

We start our investigation by looking at the switching field and the intrinsic SFD of single islands depending on the thickness of the FI-layer ($t_\mathrm{FI}=$~\SIrange{0}{50}{nm}). For each value of $t_\mathrm{FI}$, the switching fields of 100 islands with randomized microstructure and anisotropic properties, as described earlier in Section~\ref{sec:intSFD}, are computed. The curves in Fig.~\ref{fig:sfdotsize} show the average of these 100 switching fields depending on $t_\mathrm{FI}$ for three different island diameters. The standard deviation $\pm\sigma_\mathrm{int}$ for the 100 simulation runs is shown as gray shading for each curve and illustrates the intrinsic SFD.

The switching field decreases with increasing dot diameter. This is due to a change of reversal mechanism from a quasi uniform rotation Fig.~\ref{fig:Reversal}a to the nucleation and expansion of a reversed domain Fig.~\ref{fig:Reversal}e. If the thickness of the FI-layer is increased, nucleation and domain wall motion can also be seen for small dot diameters, but only in the soft layer Fig.~\ref{fig:Reversal}c. The reversal of the hard layer is still by quasi uniform rotation Fig.~\ref{fig:Reversal}d. Nonuniform reversal modes have a lower nucleation field \cite{Brown1959}. The numerical results show nonuniform reversal of the hard layer for larger diameters. In our previous work~\cite{Oezelt20152} we described this behaviour and also the influence of the exchange coupling strength at the interface on the switching field~\cite{Hauet2009}. In this work, however, we will keep the dot diameter at $d=\SI{20}{nm}$, the interface exchange coupling strength at $J_\mathrm{ixhg}=\SI{5}{mJ\per\m^2}$ and vary the thickness of the FI-layer $t_\mathrm{FI}$. 

The mean switching field $\bar{H}_\mathrm{sw}$ decreases with increasing $t_\mathrm{FI}$ as expected, but only up to a certain value ($t_\mathrm{FI}=\SI{15}{nm}$ for $d=\SI{20}{nm}$) for our set of parameters. This threshold is determined by whether the thickness of the soft ferrimagnetic layer supports the formation of a domain wall~\cite{Hagedorn1970}. We denote $t$ as thickness, $K$ as uniaxial anisotropy constant, $J$ as saturation polarization and $A$ as exchange constant of the magnetically hard ferromagnetic (FM) and the magnetically soft ferrimagnetic (FI) layer with their respective subscript. Then, for very thin soft ferrimagnetic layers the bilayers reverse their magnetization at the nucleation field~\cite{Skomski1993} of
\begin{align} H_\mathrm{n}=2\frac{t_\mathrm{FM}K_\mathrm{FM}+t_\mathrm{FI}K_\mathrm{FI}}{t_\mathrm{FM}J_\mathrm{FM}+t_\mathrm{FI}J_\mathrm{FI}}. 
\end{align}
When increasing the thickness $t_\mathrm{FI}$, a domain wall is formed in the FI-layer and gets pinned at the interface between the two layers. Therefore the switching field is now determined by the pinning field~\cite{Kronmuller2002,Suess2007}:
\begin{align}
H_\mathrm{p}=\frac{2K_\mathrm{FM}}{J_\mathrm{FM}}\frac{1-\varepsilon_K\varepsilon_A}{\left(1+\sqrt{\varepsilon_J\varepsilon_A}\right)^2},\quad 
\end{align}
where
\begin{align}
\varepsilon_K=\frac{K_\mathrm{FI}}{K_\mathrm{FM}},\quad 
\varepsilon_A=\frac{A_\mathrm{FI}}{A_\mathrm{FM}} \quad\text{and}\quad\varepsilon_J=\frac{J_\mathrm{FI}}{J_\mathrm{FM}}.
\end{align}
The required thickness $t_\mathrm{FI}$ to form a domain wall in the ferrimagnetic layer is $\delta_\mathrm{FI}=\pi\sqrt{A_\mathrm{FI}/\left(J_\mathrm{FI}H_\mathrm{p}\right)}$.

\begin{figure}[htb]
\centering
\def\svgwidth{\linewidth}
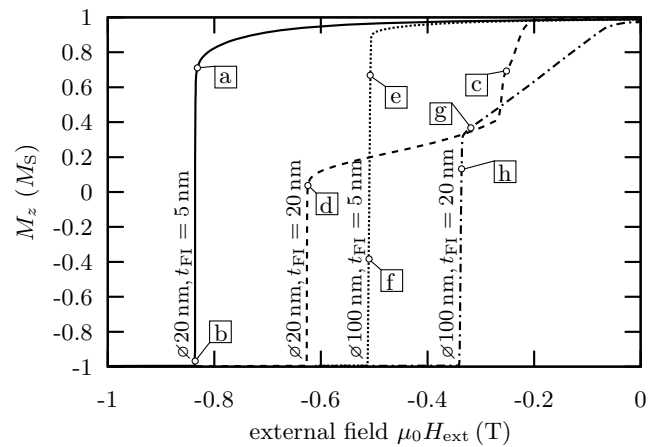
\caption{\label{fig:HystRev}Demagnetization curves of bilayer islands with diameter $d=\SI{20}{nm}$ and $\SI{100}{nm}$ and thickness of soft ferrimagnetic layer of $t_\mathrm{FI}= \SI{5}{nm}$ and $\SI{20}{nm}$. The thickness of the hard ferromagnetic layer is kept constant at $t_\mathrm{FM}=\SI{5}{nm}$. The labels a to h correspond to the magnetic states in Fig.~\ref{fig:Reversal}, two for each island design.}
\end{figure}

\begin{figure}[htb]
\centering
\def\svgwidth{\linewidth}
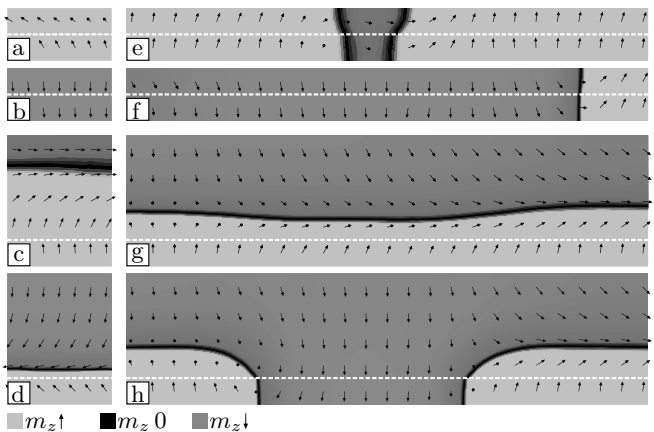
\caption{\label{fig:Reversal}Magnetic configuration during magnetization reversal of four different bilayer island designs with a $t_\mathrm{FI}$ thick ferrimagnetic layer on top and a $\SI{5}{nm}$ thick ferromagnetic layer on the bottom. The interface between the layers is the white dashed line. Left: $d=\SI{20}{nm}$ with $t_\mathrm{FI}=\SI{5}{nm}$ (a, b) and $t_\mathrm{FI}=\SI{20}{nm}$ (c, d). Right: $d=\SI{100}{nm}$ with $t_\mathrm{FI}=\SI{5}{nm}$ (e, f) and $t_\mathrm{FI}=\SI{20}{nm}$ (g, h). The labels a to h correspond to the markers in Fig.~\ref{fig:HystRev}.}
\end{figure}

In Fig.~\ref{fig:HystRev} we compare the demagnetization curves of bilayer islands with diameters $d=\SI{20}{nm}$ and $\SI{100}{nm}$ and thicknesses of the ferrimagnetic layer of $t_\mathrm{FI}= \SI{5}{nm}$ and $\SI{20}{nm}$. The thickness of the ferromagnetic layer for all four designs is $t_\mathrm{FM}= \SI{5}{nm}$. Additionally Fig.~\ref{fig:Reversal} shows two magnetic configuration states during magnetization reversal for each of the four island designs. The labels on the curves in Fig.~\ref{fig:HystRev} correspond to those in Fig.~\ref{fig:Reversal}.

When considering both, diameter and thickness of the ferrimagnetic layer, we can describe the reversal processes as follows: 
\begin{itemize}
\item small diameter / thin FI-layer (Figs.~\ref{fig:Reversal}a and~\ref{fig:Reversal}b): There is no domain wall formed, the FM-layer reverses coherently in a single step. The FI-layer reduces the switching field, but tightly follows the FM-layer.
\item small diameter / thick FI-layer (Figs.~\ref{fig:Reversal}c and~\ref{fig:Reversal}d): A domain wall is formed in the FI layer and pushed against the interface with the FM-layer. The latter switches then coherently in a single step.
\item large diameter / thin FI-layer (Figs.~\ref{fig:Reversal}e and~\ref{fig:Reversal}f): A bubble domain is formed in the FI-layer which then expands into the FM-layer. Both layers are then laterally reversed.
\item large diameter / thick FI-layer (Figs.~\ref{fig:Reversal}g and~\ref{fig:Reversal}h): A domain wall is formed in the FI-layer and pushed against the interface until the FM-layer switches laterally.
\end{itemize}

For islands with a diameter of $d=\SI{20}{nm}$, a $\SI{40}{\%}$ decrease of the mean switching field can be achieved by adding the FI-layer. At the same time this also improves the intrinsic SFD up to a thickness of around $t_\mathrm{FI}=\SI{20}{nm}$ but shows a slight increase again above this point. For the case of an island with a diameter of $d=\SI{20}{nm}$, the hard magnetic layer consists only of two grains whereas the FM layer of an island with $d=\SI{100}{nm}$ is divided in $69$ grains. This, of course, leads to a narrower SFD for the larger island, because the overall experienced anisotropy is more uniform from island to island. However, the reduction of the standard deviation due to the added FI-layer is much more pronounced for small diameters.

\begin{figure}[htb]
\centering
\def\svgwidth{\linewidth}
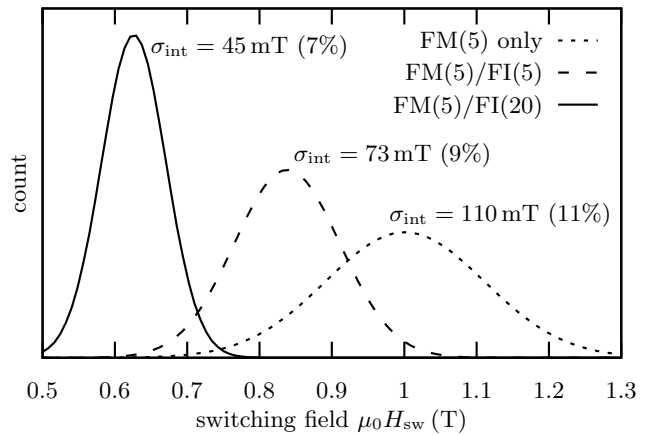
\caption{\label{fig:intrSFD}Reduction of the intrinsic SFD by coupling a FI-layer to a $\SI{5}{nm}$ thick FM-layer: Gaussian fits for single islands with no FI-layer, $\SI{5}{nm}$ and $\SI{20}{nm}$ thick FI-layer. The SFD stems solely from micro-structural variation and the variation of magnetic anisotropy properties of both layers. The standard deviation $\sigma_\mathrm{int}$ is given in absolute values and percentage of the respective mean switching field $\mu_0\bar{H}_\mathrm{sw}$.}
\end{figure}

For our further investigations we choose the smaller islands with $d=\SI{20}{nm}$, since the larger islands do not allow for a reasonable areal density of the recording medium. The variation of the FI thickness $t_\mathrm{FI}$ is reduced to only three variants: no FI-layer at all, $t_\mathrm{FI}=\SI{5}{nm}$ and $t_\mathrm{FI}=\SI{20}{nm}$. These values were chosen to compare islands without a coupled ferrimagnetic layer to islands with~\citep{Dobin2007} and without a domain wall in the soft magnetic phase during magnetic switching. We fitted the intrinsic SFD  of the islands with Gaussian distributions. In Fig.~\ref{fig:intrSFD} these fits clearly show an improvement of the intrinsic SFD when an exchange coupled ferrimagnetic layer is added. The intrinsic standard deviation is reduced by $\SI{59}{\%}$ for a FI-layer with $t_\mathrm{FI}=\SI{20}{nm}$ from $\sigma_\mathrm{int}=\SI{110}{mT}$ to $\sigma_\mathrm{int}=\SI{45}{mT}$. 

One might argue that a thicker FI-layer improves the SFD at the expense of thermal stability. We calculated the energy barrier of the various dots using the nudge elastic band method\cite{Dittrich2002}. The results show that the energy barrier for all systems is dominated by the barrier of the hard FePt (FM) layer. The calculated energy barriers are $\SI{211}{k_BT}$, $\SI{207}{k_BT}$ and $\SI{234}{k_BT}$ for FM(5), FM(5)/FI(5) and FM(5)/FI(20) respectively for $T=\SI{300}{K}$. This is in qualitative agreement with the results shown by Suess and coworkers~\cite{Suess2005}. Hard-soft bilayers islands show a higher attempt frequency than single phase islands~\cite{Dean2008, Victora2008}. With energy barriers greater than $\SI{200}{k_BT}$ even high attempt frequencies in the THz-regime give reasonable bit lifetimes.

\subsection{Dipolar interaction field contribution}
\begin{figure}[htb]
\centering
\def\svgwidth{\linewidth}
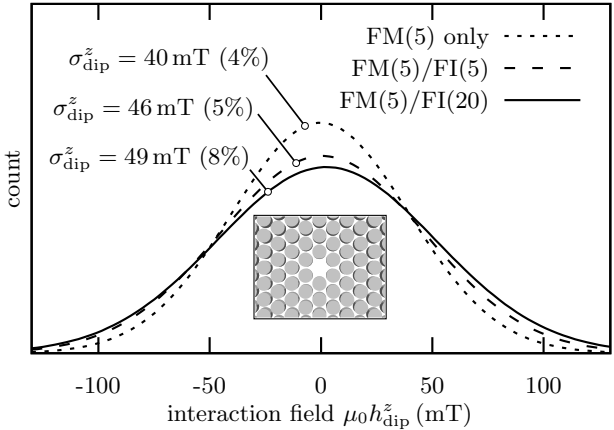
\caption{\label{fig:dip}Gaussian fits of the dipolar interaction field component $h_\mathrm{dip}^z$ distribution for $11\times11$ - arrays with triangular island pattern (inset). The distribution is shown for increasing thickness of the FI layer: FM only, $\SI{5}{nm}$ FI and $\SI{20}{nm}$ FI on top of FM. The standard deviation $\sigma_\mathrm{dip}^z$ is given in absolute values and percentage of the respective mean switching field $\mu_0\bar{H}_\mathrm{sw}$.}
\end{figure}

For the interaction field contribution to the SFD we arrange $d=\SI{20}{nm}$ islands in a triangular pattern of $11\times 11$ bits with a virtual island in the center where the probe point is located (see inset of Fig.~\ref{fig:dip}). Their microstructure and the anisotropic variation are stripped away and all easy axes are set perfectly out-of-plane. The distribution of the interaction field acting on an island is computed as described in Section~\ref{sec:dipSFD}. Each distribution is compiled from the result of 500 simulation runs with random initial magnetization. Fig.~\ref{fig:dip} depicts the Gaussian fits of the out-of-plane component of the interaction field distributions $h_\mathrm{dip}^z$. 

\begin{figure}[htb]
\centering
\def\svgwidth{\linewidth}
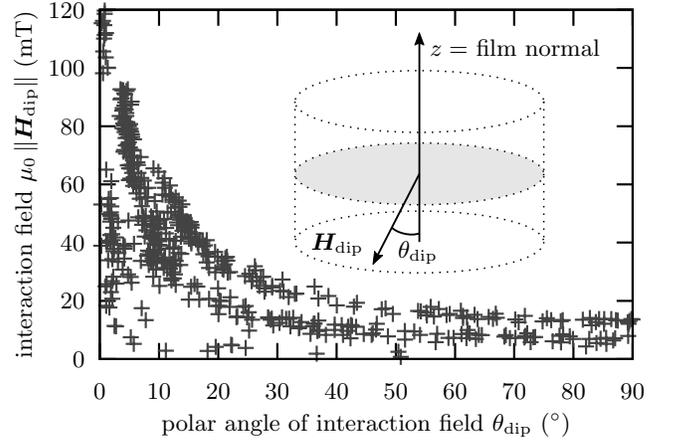
\caption{\label{fig:avh5}Magnitude of the interaction field plotted against the respective polar angle $\theta_\mathrm{dip}$ for 500 simulation runs of FM($\SI{5}{nm}$)/FI($\SI{5}{nm}$) island arrays with random initial magnetization.}
\end{figure}

In Fig.~\ref{fig:avh5} we show how the angle $\theta_\mathrm{dip}$ between the film normal and the interaction field (see inset) is distributed. We computed the interaction field of 500 random magnetization configurations of the bits in the array and plotted the magnitude of the field as function of the field angle. In general, higher interaction fields can be observed at smaller angles. The results indicate, that in a configuration that leads to an interaction field with a high angle, the field is small. Further we see that most points in Fig.~\ref{fig:avh5} are located in the range from $\SI{0}{\degree}$ to $\SI{20}{\degree}$.

For the calculation of the BER later on, all three orthogonal components of $\vect{H}_\mathrm{dip}$ are considered. Following \eqref{eq:hteff} we compute the distribution of the total field acting on the island from the 500 samples.
We again look at islands of single hard FM-layer and bilayer islands with exchange coupled $\SI{5}{nm}$ and $\SI{20}{nm}$ thick soft FI-layers. When adding the FI-layer an increase in $\sigma_\mathrm{dip}$ of up to $\SI{23}{\%}$ ($\SI{9}{mT}$) for $t_\mathrm{FI}=\SI{20}{nm}$ is observed. This can be attributed to the increased height of the magnetic islands when the FI-layer is added. This change of the island shape increases the stray field acting on a neighbouring island. To test the influence of the number of considered neighbouring islands we did the same calculations for a $17\times17$ - array. In arrays of bilayer islands with $t_\mathrm{FI}=\SI{20}{nm}$, the maximum deviation was found to be $\SI{0.55}{mT}$ for $\sigma_\mathrm{dip}^z$. Therefore we think the choice of $11\times11$ - arrays is justified.

\subsection{\label{sec:berRes}Bit error rate}
\begin{figure}[htb]
\centering
\def\svgwidth{\linewidth}
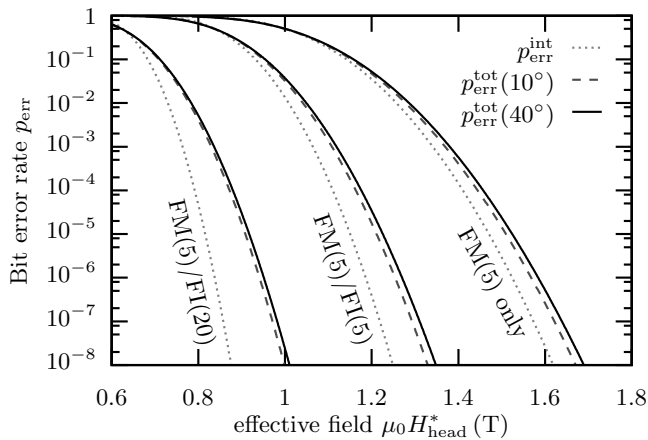
\caption{\label{fig:bercomp}Calculated bit error rates for single layer islands with no FI-layer and for bilayer islands with $\SI{5}{nm}$ and $\SI{20}{nm}$-thick exchange coupled FI-layer in $\SI{1.4}{Tb/in^2}$ triangular patterned BPM. For all three configurations, the BER without dipolar field contribution $\ber{int}$ is shown as dotted line. The BER incorporating the diploar field contribution $\ber{tot}$ in addition is calculated at $\varphi=\SI{10}{\degree}$ (dashed line) and $\varphi=\SI{40}{\degree}$ (solid line) head field angle for the three designs.}
\end{figure}

Generally it can be stated, that adding the FI-layer narrows the intrinsic SFD but broadens the distribution of the interaction field acting on an island. As described in section~\ref{sec:ber} we now calculate the BER with an effective field incorporating the angular variation of the switching field. We do this with and without the dipolar interaction contribution to look at its influence on the BER. The calculations are again done for the three island configurations: single FM-layer with $\SI{5}{nm}$ thickness and exchange coupled bilayers with $\SI{5}{nm}$ and $\SI{20}{nm}$ FI-layer on top of the FM-layer. For this comparison we choose to apply the writing field $\vect{H}_\mathrm{head}$ at two angles $\varphi=\SI{10}{\degree}$ and $\varphi=\SI{40}{\degree}$. In Fig.~\ref{fig:bercomp} the bit error rates are plotted against the effective head field $H_\mathrm{head}^*$. 

As shown in section~\ref{sec:intSFD}, the switching field is decreased when adding the exchange coupled FI-layer. Therefore we see a strong reduction of the BER for a given $H_\mathrm{head}^*$ with increasing $t_\mathrm{FI}$, regardless of the method used to calculate $\ber{}$. Comparing the two methods $\ber{int}$ and $\ber{tot}$ we immediately see that incorporating the dipolar interaction field can have a significant impact. The discrepancy between both methods increases when adding an exchange coupled FI-layer because $\sigma_\mathrm{int}$ decreases while $\sigma_\mathrm{dip}$, and consequently $\sigma_\mathrm{tot}$, increases slightly and contributes much more. For example, the FM(5)/FI(20) island design at an effective head field of $\mu_0H_\mathrm{head}^*=\SI{838}{mT}$ with $\varphi=\SI{40}{\degree}$ shows a BER of $\ber{int}=10^{-6}$ without a dipolar field contribution, but a much higher BER of $\ber{tot}=10^{-3}$ when incorporating the interaction field.
For $\ber{int}$, a change in $\varphi$ only changes $H_\mathrm{head}^*$ and therefore does not alter the curves in Fig.~\ref{fig:bercomp}. If the dipolar field is taken into account there is only a neglectable difference for different head field angles.

In TABLE~\ref{tab:ber}, the standard deviation for the intrinsic and dipolar contribution to the SFD as well as the switching fields are listed for the three island designs. We also show the respective minimum required effective head field for a given BER of $\ber{}=10^{-6}$, with and without incorporating $\vect{H}_\mathrm{dip}$. From this point of view, if we neglect the interaction field, the required head field for $\ber{}=10^{-6}$ and $\varphi=\SI{40}{\degree}$ is reduced from $\SI{1524}{mT}$ to $\SI{838}{mT}$ when a $\SI{20}{nm}$ FI-layer is added. If we also consider the dipolar field as we proposed, the required field is $\SI{951}{mT}$. In other words: When taking the dipolar field into account, we still see a significant reduction of the required writing field, but only by $38\%$ as opposed to the $45\%$ we gain when neglecting $\vect{H}_\mathrm{dip}$.
\begin{table}[htp]
\centering
\caption{\label{tab:ber}Switching field distribution and bit error rate for the three island designs: FM(5) only, FM(5)/FI(5) and FM(5)/FI(20) bilayer. The required effective head fields $H_\mathrm{head}^*$ are listed for a BER of $\ber{}=10^{-6}$ for both calculation methods and two different head field angles $\varphi=10$ and $\SI{40}{\degree}$.}
\begin{ruledtabular}
\begin{tabular}{@{\extracolsep{\fill}}lllcrrr}
 & & FM(5)/FI($t_\mathrm{FI}$) & (nm) & \multicolumn{1}{c}{0} & \multicolumn{1}{c}{5} & \multicolumn{1}{c}{20} \\
\hline
 & & $\sigma_\mathrm{int}$ & (mT) & 110 & 73 & 45 \\
 & & $\sigma_\mathrm{dip}^{z}$ & (mT) & 40 & 46 & 49 \\
 & & $\mu_0\bar{H}_\mathrm{sw}$ & (mT) & 1001 & 838 & 626\\
 \hline
\multirow{5}{*}{\rotatebox[origin=c]{90}{$\ber{}=10^{-6}$}} & & $\vphantom{3^{3^{3^3}}} \mu_0H_\mathrm{head}^*(\ber{int})$ & (mT) & 1524 & 1186 & 838\\
 \cline{2-7}
 & \multirow{2}{*}{\rotatebox[origin=c]{90}{\SI{10}{\degree}\footnotemark[1]}} & $\sigma_\mathrm{tot}$ & (mT) & 46 & 48 & 50\\
 & & $\mu_0H_\mathrm{head}^*(\ber{tot})$ & (mT) & 1568 & 1254 & 941\\
 \cline{2-7}
 & \multirow{2}{*}{\rotatebox[origin=c]{90}{\SI{40}{\degree}\footnotemark[1]}} & $\sigma_\mathrm{tot}$ & (mT) & 54 & 54 & 53\\
 & & $\mu_0H_\mathrm{head}^*(\ber{tot})$ & (mT) & 1585 & 1271 & 951
\end{tabular}
\end{ruledtabular}
\footnotetext[1]{Angle $\varphi$ between applied head field and out-of-plane axis.}
\end{table}

\subsection{Read back field}
\begin{figure}[htb]
\centering
\def\svgwidth{\linewidth}
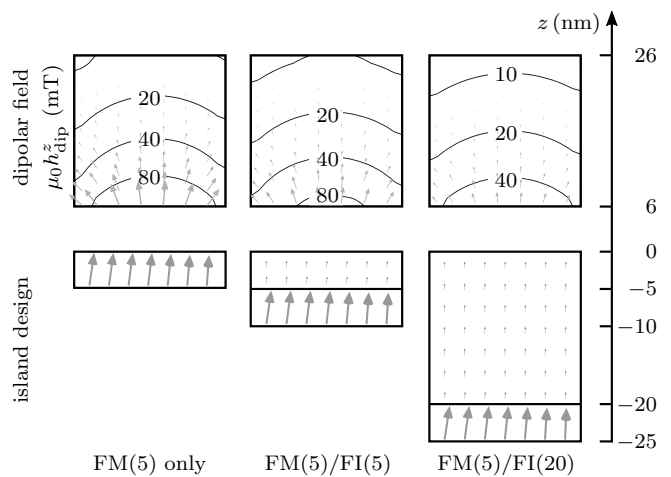
\caption{\label{fig:readback} Calculated stray field (upper boxes) of three single island designs (lower boxes) with a diameter of $d=\SI{20}{nm}$ experienced by a read head with a magnetic spacing of $\SI{6}{nm}$. The magnetic moments in the island layers and the stray field above the islands are shown as gray arrows. The isolines in the field boxes show the stray field in mT.}
\end{figure}

We computed the field above the island at the place where the read head would be positioned. We assumed a magnetic spacing of $\SI{6}{nm}$. A map of the perpendicular magnetic field component is shown in Fig.~\ref{fig:readback}. As expected the field decreases with increasing ferrimagnetic layer thickness. This is mostly due to low magnetization of FeGd. The figure also shows the magnetization configurations in the two layers and the vectors of the magnetostatic field. The perpendicular field components $\mu_0h^z_\mathrm{dip}$ are in the range from $\SI{10}{mT}$ to $\SI{40}{mT}$ for the thickest ferrimagnetic layer.

\section{Summary}
In this work we described the micromagnetic models to compute the intrinsic contribution and the dipolar interaction field contribution to the switching field distribution (SFD) and switching field of bit patterned media of ferri-/ferromagnetic bilayer islands. A method is proposed to compute the bit error rate (BER) incorporating both contributions and also the angle of the writing field. 
A decrease of the island diameter (to increase the areal density) changes the reversal mechanism into the single domain regime, and therefore increases the switching field and broadens the switching field distribution.
Our results show that adding an exchange coupled ferrimagnetic (FI) soft layer decreases the switching field and its distribution significantly. This also greatly reduces the bit error rate of bit patterned media. Both the switching field and its distribution decrease with increasing thickness of the ferrimagnetic soft layer, especially up to a thickness which then allows the formation of a domain wall. But with increasing thickness of FI layer and increasing areal density the influence of the dipolar interaction field becomes increasingly important. We conclude that, up to a certain thickness of the soft FI-layer, the bilayer island design is very beneficial for the bit error rate of bit patterned media, but the dipolar field has to be taken into consideration. 

In heat assisted magnetic recording bits are addressed for writing by the intersection of the magnetic write field with the heat spot. By careful design of the magnetic write field profile single islands for our chosen dimensions can be addressed~\cite{Kovacs2016}. With regard to heat assisted magnetic recording, the exchange coupled ferrimagnetic soft layer should be designed to be at its compensation point, exerting no dipolar interaction field. Only bilayer islands applied with the laser heat spot will have a reduced switching field and will only experience the interaction field of the neighbouring hard phase of the islands. By using this scheme, the advantage of exchange spring media can be exploited while keeping the magnetostatic interaction field low. The method proposed in this work can be used to find the optimal thickness of the FI-layer for certain parameters of head field, head field angle and a desired bit error rate.

\begin{acknowledgments}
We gratefully acknowledge the financial support provided by the Austrian Science Fund (FWF Grant No. I821), the German Research Foundation (DFG Grant No. AL 618/17-1), the Swiss National Science Foundation (SNF Grant No. 200021L\_137509) and the Vienna Science and Technology Fund (WWTF Grant No. MA14-044).
\end{acknowledgments}

\bibliography{references}

\end{document}